%% file: ms.tex
\documentclass[12pt,preprint]{aastex}

\received{2001 October 31}
\begin{document}

\title{SDSS~J124602.54+011318.8: A Highly Luminous
  Optical Transient at $z=0.385$\altaffilmark{1}}

\author{
Daniel E. Vanden Berk\altaffilmark{2},
Brian C. Lee\altaffilmark{2},
Brian C. Wilhite\altaffilmark{2,3},
John F. Beacom\altaffilmark{2},
Donald Q. Lamb\altaffilmark{3},
James Annis\altaffilmark{2},
Kevork Abazajian\altaffilmark{2},
Timothy A. McKay\altaffilmark{4},
Richard G. Kron\altaffilmark{2,3},
Stephen Kent\altaffilmark{2,3},
Kevin Hurley\altaffilmark{5},
Robert Kehoe\altaffilmark{4,6},
Jim Wren\altaffilmark{7},
Arne A. Henden\altaffilmark{8,9},
Donald G. York\altaffilmark{3,10},
Donald P. Schneider\altaffilmark{11},
Jennifer Adelman\altaffilmark{2},
Jon Brinkmann\altaffilmark{12},
Robert J. Brunner\altaffilmark{13},
Istv\'an Csabai\altaffilmark{14,15},
Michael Harvanek\altaffilmark{12},
Greg S. Hennessy\altaffilmark{9},
\v{Z}eljko Ivezi\'{c}\altaffilmark{16},
Atsuko N. Kleinman\altaffilmark{12},
Scot J. Kleinman\altaffilmark{12},
Jurek Krzesinski\altaffilmark{12,17},
Daniel C. Long\altaffilmark{12},
Eric H. Neilsen\altaffilmark{14},
Peter R. Newman\altaffilmark{12},
Stephanie A. Snedden\altaffilmark{12},
Chris Stoughton\altaffilmark{2},
Douglas L. Tucker\altaffilmark{2},
Brian Yanny\altaffilmark{2}
}

\altaffiltext{1}{Based on observations obtained with the Sloan Digital
  Sky Survey, which is owned and operated by the Astrophysical Research
  Consortium.}
\altaffiltext{2}{Fermi National Accelerator Laboratory, P.O. Box 500,
  Batavia, IL 60510}
\altaffiltext{3}{The University of Chicago, Department of Astronomy
  and Astrophysics, 5640 S. Ellis Ave., Chicago, IL 60637}
\altaffiltext{4}{University of Michigan, Department of Physics,
  500 East University, Ann Arbor, MI 48109}
\altaffiltext{5}{ University of California at Berkeley,
  Space Science Laboratory, Grizzly Peak and Centennial Drive, Berkeley,
  CA 94720}
\altaffiltext{6}{Michigan State University, Department of
  Physics and Astronomy, East Lansing, Michigan 48824}
\altaffiltext{7}{Los Alamos National Laboratory, PO Box 1663, Los Alamos,
  NM 87545}
\altaffiltext{8}{Universities Space Research Association}
\altaffiltext{9}{U.S. Naval Observatory, 3450 Massachusetts Ave.,
  NW, Washington, DC  20392-5420}
\altaffiltext{10}{The University of Chicago, Enrico Fermi Institute,
  5640 S. Ellis Ave., Chicago, IL 60637}
\altaffiltext{11}{Department of Astronomy and Astrophysics,
  The Pennsylvania State University, University Park, PA 16802}
\altaffiltext{12}{Apache Point Observatory, P.O. Box 59, Sunspot,
  NM 88349-0059}
\altaffiltext{13}{Department of Astronomy, California Institute of
  Technology, Pasadena, CA 91125}
\altaffiltext{14}{Department of Physics and Astronomy,
  The Johns Hopkins University, 3701 San Martin Drive, Baltimore, MD 21218}
\altaffiltext{15}{Department of Physics of Complex Systems, E\"otv\"os
  University, P\'azm\'any P\'eter s\'et\'any 1}
\altaffiltext{16}{Princeton University Observatory, Princeton, NJ 08544}
\altaffiltext{17}{Mt.\ Suhora Observatory, Cracow Pedagogical
  University, ul.\ Podchorazych 2, 30-084 Cracow, Poland}

\begin{abstract}
We report the discovery of a highly luminous optical transient (OT),
SDSS~J124602.54+011318.8, associated with a galaxy at a redshift
of 0.385.  In this paper we consider the possibility that the OT may
be a GRB afterglow. Three sets of images and two sets of spectra were
obtained as part of the normal operations of the Sloan Digital Sky
Survey (SDSS).  In the first two image sets, observed two nights
apart, the object appears as a point source at $r^{*}\approx 17$.
The third image set, observed about 410 days later, shows an extended
source which is more than $2.5$ magnitudes fainter.  The spectra
were observed about 400 and 670 days after the first two image sets,
and both show an apparently normal galaxy at a redshift of 0.385.
Associating the OT with the galaxy, the absolute magnitude was
$M_{r^*}=-24.8$, which is over 4 magnitudes brighter than the most
luminous supernova ever measured.  The spectral energy distributions
of the galaxy-subtracted OT derived from the first two image sets
are well-fit by single power-laws with indices of $\beta_{\nu}=-0.92$
and $-1.29$ respectively, similar to most GRB afterglows.  Based upon
the luminosity of the OT, non-detections in contemporaneous ROTSE-I
images, and the change in spectral slope, the OT, if an afterglow,
was likely discovered early during a ``plateau'' or slowly-fading
phase.  The discovery of a GRB afterglow at this stage of the SDSS
is consistent with expectations, but only if the optical emission
is much less strongly beamed than the gamma-rays.  We emphasize that
other explanations for the OT cannot be ruled out; a recent follow-up
study by \citet{galyam02} provides strong evidence that this source
is in fact an unusual AGN.
\end{abstract}


\section{Introduction}
All gamma-ray burst (GRB) afterglows discovered to date have
been found by follow-up optical or radio imaging of satellite gamma-ray
detection localizations. The optical follow-up observations have
helped to identify GRB host galaxies, and to determine redshifts,
spectral-energy distributions (SED), and decay rates for the optical
counterparts.  The GRB afterglows observed so far typically exhibit a
power-law SED of about $f_\nu \propto \nu^{-1}$, and a luminosity
time decay which is also usually well-fit by one or two power-laws
with indices of $-0.7$ to $-2.4$ \citep{stanek99}.  The lack of detections 
of optical
GRB afterglows in supernova searches and other variable object
surveys has placed constraints on the relative gamma-ray to optical
beaming angles, and hence on the GRB rate and energy output
\citep{rhoads97}.

The Sloan Digital Sky Survey \citep[][SDSS]{york00} was not
designed to search for GRB afterglows, or any other transient
objects.  The explicit use of the camera for such follow-up can
only occur under rare favorable conditions \citep[e.g.,][]{lee01}.
However, many classes of highly variable or transient objects
have photometric colors which separate them from the locus of main
sequence stars in the SDSS color space.  Therefore such objects
may, in principle, be targeted for spectroscopic follow-up in the
SDSS \citep[c.f.][]{vandenberk02,stoughton02,rhoads01,krisciunas98}.
We show here that at least one object with properties similar to known
GRB afterglows has been detected in the SDSS imaging survey, with
follow-up observations in the spectroscopic survey.  If the object is
indeed a GRB afterglow, then it is the first to be found entirely
with optical methods.

\section{Observations and Selection Techniques \label{dataset}}
The SDSS is a project to image $10^{4}~{\rm deg}^{2}$
of sky centered roughly on the Northern Galactic Pole, in five
broad photometric bands ($u,g,r,i,z$), to a depth of 
$r\sim 23$, and to obtain spectra for $10^6$ galaxies and $10^5$ 
quasars selected from the imaging survey.  The observations are
made with a dedicated 2.5~m telescope \citep{siegmund02}
equipped with a large mosaic CCD camera \citep{gunn98} and
a pair of fiber-fed spectrographs \citep{uomoto02}. Technical details of 
the survey data acquisition system and reduction are given by
\citet{york00} and \citet{stoughton02}.

\begin{figure}
\includegraphics{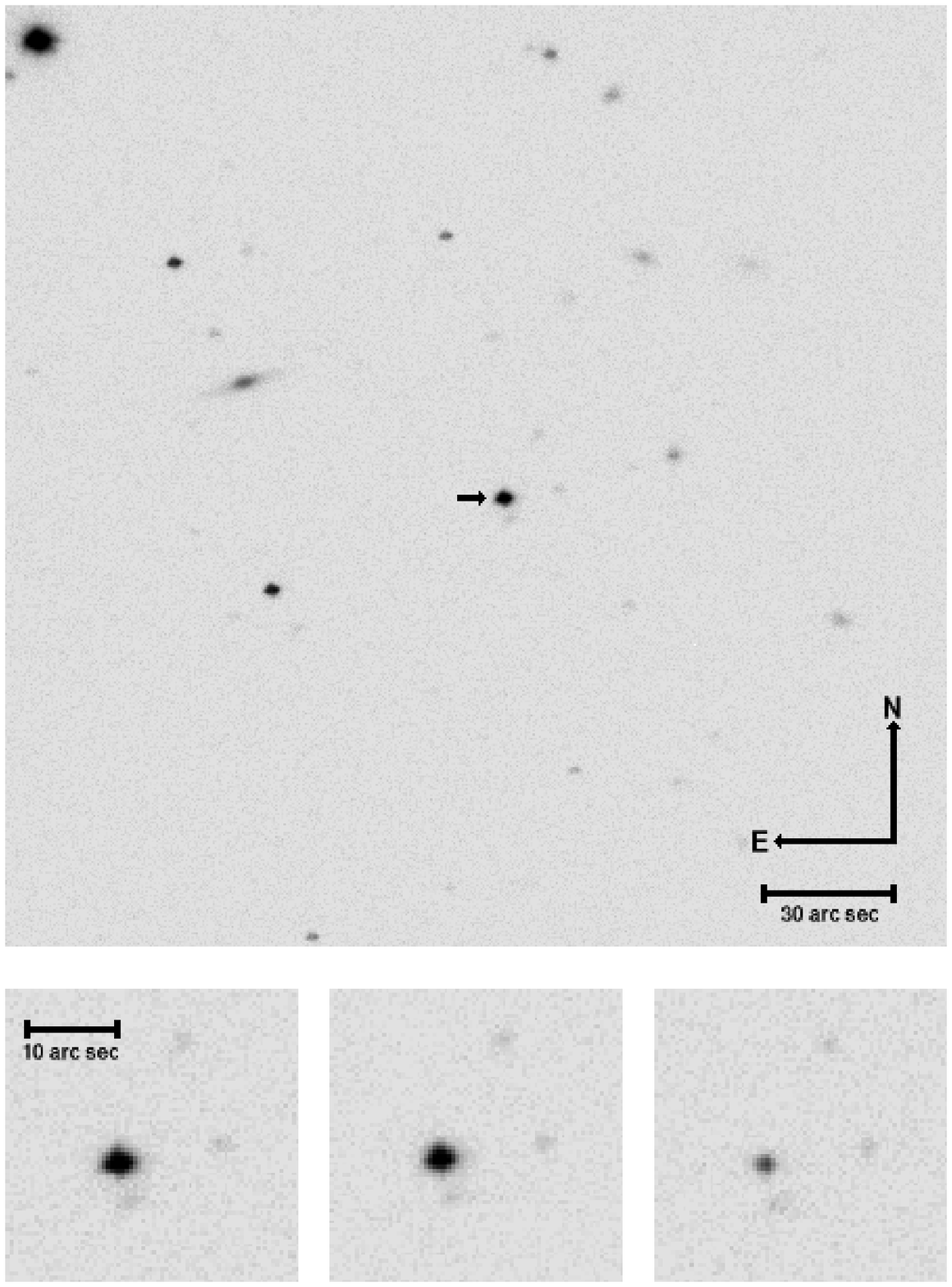}
\caption{ Finder chart $r$ band image of a $3\arcmin\times3\arcmin$ field
containing the optical transient, SDSS~J124602.54+011318.8, from run 745.
Also shown are three $30\arcsec\times30\arcsec$ $r$ band images
from run 745 (1999 March 20), 756 (1999 March 22), and 1462 (2000 May 5)
respectively.  The first two images are consistent with the point spread 
function, while the third is slightly extended.
\label{image}}
\end{figure}

Images are taken nearly simultaneously through 5 filters,
$ugriz$, reduced with the SDSS photometric reduction pipeline
PHOTO \citep{lupton01}, and calibrated in the preliminary SDSS
photometric system, $u^*g^*r^*i^*z^*$, which may differ by at
most a few percent from that described by \citet{fukugita96}
\citep[c.f.][]{stoughton02}.  Three sets of images each containing the
coordinates of the optical transient were taken on 1999 March 20,
1999 March 22, and 2000 May 5, denoted by run numbers 745, 756, and
1462 respectively. 
The OT was observed at $12\rm{h} 46\rm{m} 02\fs54, +1\arcdeg 13\arcmin 
18\farcs8$.
The magnitudes and coordinates on each of the
three nights are given in Table\,\ref{obstable}. A finder chart and
comparison images from each night are shown in Figure\,\ref{image}.
The profiles of the object on the first two nights are consistent with
the point spread function (PSF) of the images (see \citet{lupton01}
and \citet{stoughton02} for a discussion of the SDSS PSF).
The object has an $r$
band half-light radius of $0\farcs79$ in the third image set, and is
extended relative to the PSF, indicative of a galaxy.  The relative
astrometric accuracy across runs is $0\farcs07$ in both RA and DEC,
and all three coordinates are consistent within this uncertainty.

\begin{figure}
\includegraphics[scale=0.8]{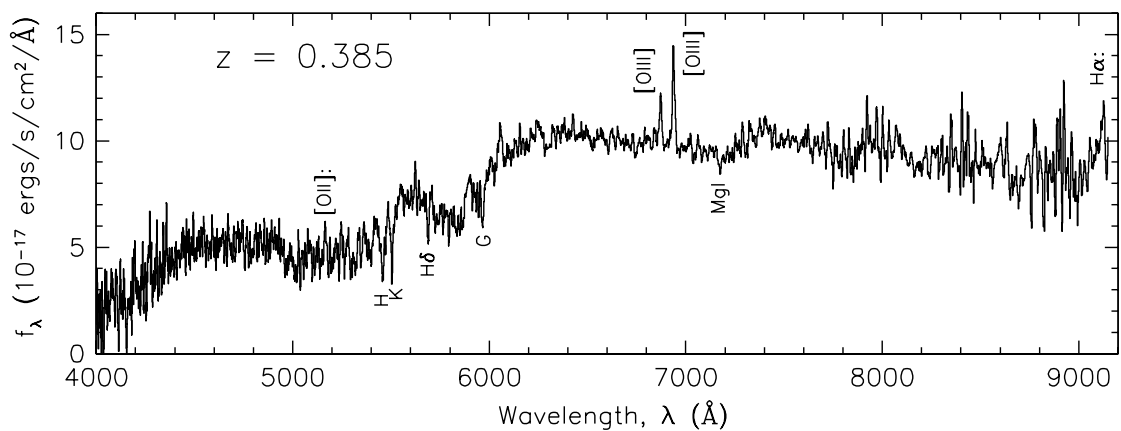}
\caption{Coadded SDSS spectrum of the galaxy at the OT coordinates.
The spectrum has been smoothed by 5 bins to improve the signal-to-noise
ratio, and truncated below $4000${\AA} to omit the region of highest
noise.  The spectrum has been corrected for telluric absorption 
\citep[see][]{stoughton02}.  The 
smoothed resolution is about 800.  Several emission
and absorption lines are labeled. The [O\,{\sc ii}] and H\,$\alpha$
lines are uncertain.  The [Ne\,{\sc iii}]$\lambda3869$ emission line 
sometimes seen in the spectra of GRB hosts and AGN is not present in this 
spectrum.
\label{spectrum}}
\end{figure}

The OT was selected for spectroscopic follow-up as a quasar candidate
based upon color criteria and a FIRST radio source optical match, from
the imaging data in run 756.  Several types of transient or otherwise
variable objects -- including cataclysmic variables, supernovae,
and GRB afterglows --  may be selected by the quasar algorithms,
since they occupy some of the same volume of color space in the SDSS
filter system as quasars \citep{vandenberk02,rhoads01,krisciunas98}.
The OT was targeted on the SDSS spectroscopic plate number 291, and
spectra were obtained on 2000 April 26 and 2001 January 19.  The
spectra were reduced and calibrated using the SDSS spectroscopic
pipeline \citep{frieman02}.  The coadded spectrum is shown in
Figure\,\ref{spectrum}, and reveals a galaxy at a redshift of
$z=0.385$.

The OT was discovered as part of a program to find
supernovae and other highly variable objects in the SDSS dataset,
details of which are given by \citet{vandenberk02}.  Briefly, the
flux-calibrated spectra are convolved with the SDSS $g,r$, and $i$
filter transmission curves (the $u$ and $z$ bands are not fully covered
by the spectra), including 1.2 airmasses of extinction and the CCD
response function, to generate synthetic spectroscopic magnitudes.
Spectroscopic targets which are significantly fainter in the spectra
compared to the images are flagged as variable object candidates.
Each candidate's spectrum and corresponding image are examined in order
to identify the object type.  The optical transient faded by $\approx
2.5$ magnitudes in all three bands, and the normal galaxy spectrum
made it a candidate supernova or GRB afterglow.  The spectroscopic
magnitudes are given in Table\,\ref{obstable}.

\section{Results\label{results}}

\begin{figure}
\includegraphics[scale=0.7,angle=-90]{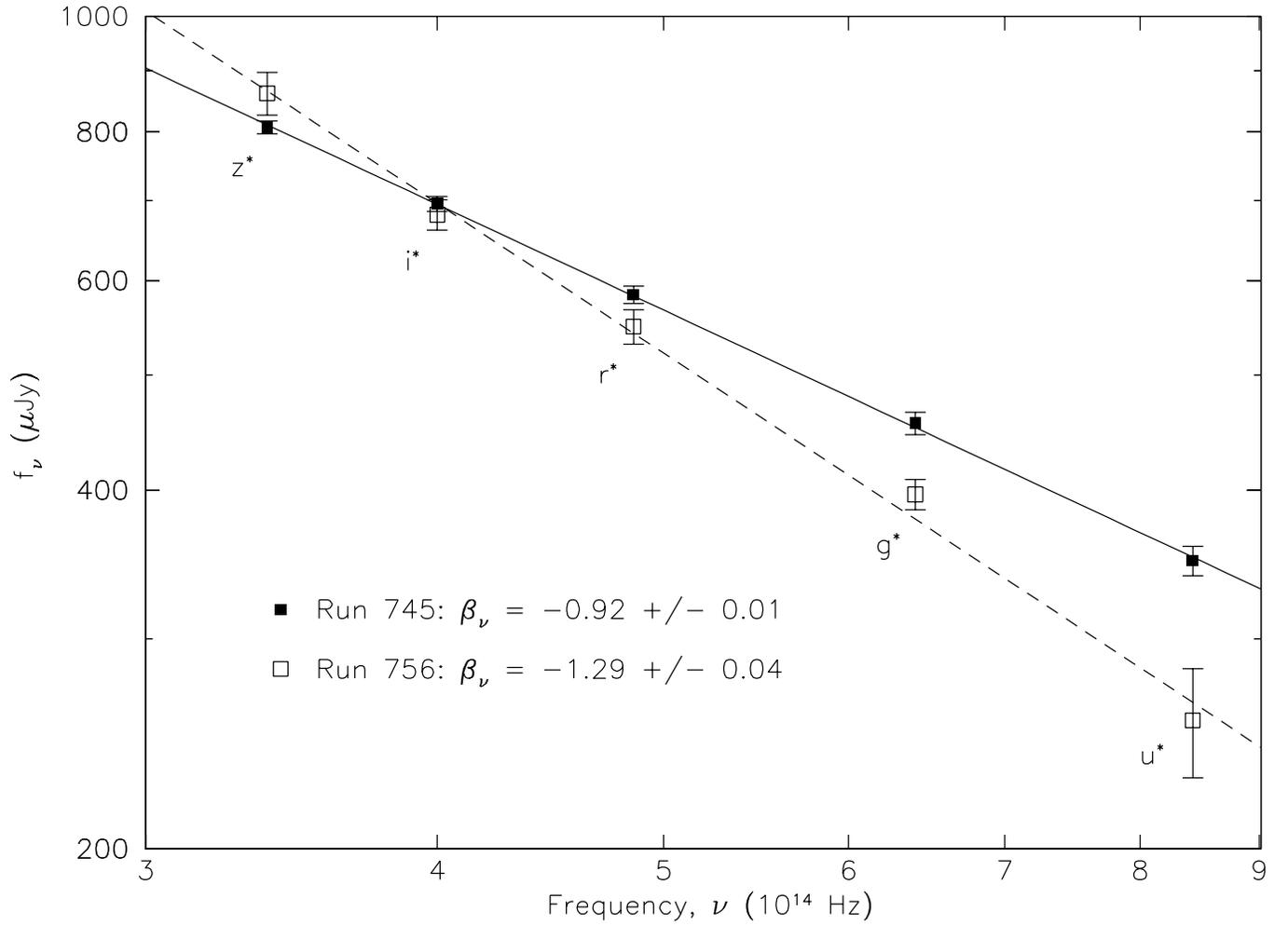}
\caption{Spectral energy distribution of the OT in the observer frame,
with the galaxy light removed, and corrected for Galactic reddening, from
imaging runs 745 and 756.  Power law fits to the SEDs are shown and the
indices are given.  The errorbars include magnitude uncertainties in both
the OT and underlying galaxy.
\label{sed}}
\end{figure}

We have generated spectral energy distributions for the
OT in its bright phase by transforming the reddening-corrected
\citep{schlegel98} and galaxy-subtracted magnitudes into flux
density values according to the $AB_\nu$ system as described by
\citet{fukugita96}.  The SEDs of the OT light in runs 745 and 756
are shown in Figure\,\ref{sed}.  Both SEDs are well-fit by power
laws of the form $f_{\nu} = (\nu / \nu_0)^{\beta_{\nu}}$, with index
values $\beta_{\nu} = -0.92\pm0.01$ (run 745) and $-1.29\pm0.04$
(run 756).  Both values fall well within the typical range for GRB
afterglows ($-2.3 \la \beta_{\nu} \la -0.7$).  If the OT is at the
same redshift as the galaxy, $z=0.385$, and using a flat
$\Lambda$-dominated cosmology ($\Omega_m=0.3, \Omega_\Lambda=0.7,
H_0=65~{\rm km~s^{-1}~Mpc^{-1}}$), the absolute magnitudes
of the OT on 1999 March 20 and 22 were $M_{r^*}=-24.75$ and
$M_{r^*}=-24.81$ respectively.
This is about 6 magnitudes brighter than Type~Ia supernovae at
maximum \citep[c.f.][]{vaughan95}, and as much as 4 magnitudes
brighter than the most luminous supernova ever discovered (SN~1997cy, 
which was discovered perhaps a few weeks post-maximum
\citep{germany00}).  In addition, the colors of the OT are not 
similar to those of the most commonly classified supernovae
\citep[although we cannot rule out some rare and unusual class
of supernova by colors alone]{poznanski02}.

Before discussing the implications of the GRB afterglow hypothesis,
we consider other potential identifications.  The OT is unlikely to
be within the solar system as Earth reflex motion would cause an
offset of greater than $0\farcs3$ ($3\sigma$ detection) between
the $g$ and $r$ image positions for objects out to the Kuiper
Belt \citep{ivezic00}.  The only known types of variable stars that
cannot be excluded based upon insufficient variability amplitude or
non-matching colors and SEDs are some types of cataclysmic variables.
We have examined 22 cataclysmic variables with SDSS photometry
\citep{szkody02,krisciunas98} and find only 2 or 3 which are reasonably
consistent with power-law SEDs. 
There is also no evidence for the spectral signature of any type
of galactic variable star.  In addition, we find no evidence for past
variability from a visual examination of the publicly available images
extracted from the Digitized Sky Survey\footnote{The Digitized Sky
Survey was produced at the Space Telescope Science Institute under U.S.
Government grant NAGW-2166; the digitized image data is publicly
available, but the magnitude-calibrated data is not.} of photographic
plates observed in 1956, 1979, 1990, and 1991.  Furthermore, the
remarkably close coincidence with a galaxy would have to be explained
as a random alignment (for example, the probability of a random
coordinate falling within 0.5\arcmin of a galaxy brighter than $r^* =
20.0$ is $\approx 10^{-4}$ \citep{yasuda01}).  Based upon these
arguments, it is unlikely that the OT can be identified with any type
of Galactic object.

Aside from GRB afterglows, the only known types of variable objects
luminous enough to identify with the OT are quasars and other active
galactic nuclei (AGN).  The SEDs of AGN can often be approximated by
power laws with indices $\beta_{\nu}\sim -0.5$.
Also, the host galaxy is coincident with a radio source detected
in the FIRST survey \citep{becker95} with integrated flux $S =
79.4$~mJy at 1.4~GHz, and the PMN survey \citep{griffith95} with
$S = 51$~mJy at 4.9~GHz.  The radio flux is high for a ``normal''
galaxy \citep{magliocchetti00}, and would require a star formation
rate $\gtrsim 30,000 M_{\sun}$ per year to explain it \citep[see,
e.g.][]{condon92,berger01}.  However, the galaxy is not detected in
the far infrared in the IRAS catalogs \citep{beichman88}.  These facts
imply that AGN activity rather than star formation is the source of
the radio emission.  Therefore the galaxy probably harbors an AGN
which is obscured, perhaps entirely, at optical wavelengths.
However, the defining optical characteristics of an AGN -- a
non-stellar spectrum with strong broad emission lines -- are not seen
in the galaxy spectrum.  The galaxy is also undetected in the ROSAT
All Sky Survey \citep{voges99}, and a search of all other available
X-ray surveys using the NASA High Energy Astrophysics Science Archive
Research Center reveals no known X-ray source.  This nearly precludes
the identification of the galaxy as a blazar -- a class of AGN with
strong X-ray emission which have been observed to vary at optical
wavelengths by several magnitudes on timescales of months to years
\citep[e.g.][]{schramm94,webb88}.
Also, as described above, we found no evidence for past optical
variability.  The coordinates set the location of the OT at
$0.5\pm0.5$~kpc from the galaxy center, which is consistent with
other known GRB afterglows \citep{bloom02}.  These facts suggest that
although the galaxy probably harbors an obscured AGN, the OT is not
the result of normal AGN activity.

If the galaxy contains an AGN, an intriguing possibility for the OT
is that of a tidally disrupted star near a supermassive black hole
\citep{rees88}, which may cause a bright relatively short-lived flare.
However, estimates of the spectral signature of such events predict
either thermal emission from the accretion disk at $\sim10^{5}$K
\citep{ulmer99}, which would be too blue to match the OT, or
reprocessing of the radiation by tidal debris at $\sim10^{4}$K lasting
at least several years \citep{loeb97}, which also would not match the
observed properties of the OT.  In addition, any flare associated with
an obscured AGN would still have to overcome the optical obscuration.
While we find the evidence for an AGN less consistent than for a GRB
afterglow, we cannot completely rule out the possibility that the
OT is some kind of extreme AGN.  Further observations such as higher
signal-to-noise ratio spectra, high resolution radio mapping, 
and especially long-term optical monitoring (AGN activity may
recur while an afterglow will not) would place further
constraints on either the afterglow or AGN hypothesis.

Another possibility is that the object is an AGN of some sort which is 
reddened to have the observed SEDs.  For example, using the reddening function
given by \citet{francis00}, one can transform a pure $\beta_{\nu}=-0.3$ 
power-law into an SED that roughly approximates a power-law with index -1 in 
the optical, using $E(B-V)=0.15$.  However, the resulting SED is only poorly
matched by a power-law, and to achieve a slope in the correct range requires
a fine-tuning of the color excess.  Similar attempts to transform a hot 
thermal accretion disk SED into the appropriate power-law fail for all values
of the color excess.  In addition, although observed AGN power-law slopes can 
be as red as this OT, values of $\beta_{\nu}\approx -1$ are well out on the 
tail of the distribution \citep[e.g.][]{francis96}.

If the OT is a GRB afterglow, then the properties of its presumed
host galaxy are of interest.  The absolute magnitude of the galaxy (in
run 1462) at $z=0.385$ is $M_{r^*} = -22.3$ -- slightly brighter than
an $L_{*}$ galaxy in the SDSS photometric system \citep{blanton01},
and near the median value for GRB host galaxies \citep{djorgovski01}.
Based upon the possible detection of narrow [O\,{\sc ii}]$\lambda
5007$ and H\,$\alpha$ emission lines, we estimate the star
formation rate to be $SFR \approx 1{\rm M}_\sun~{\rm yr}^{-1}$,
\citep{kennicutt98}.
We have argued above that the radio flux is most likely due to
an AGN that is heavily obscured at optical wavelengths.  The star
formation rate estimated from the emission line flux is therefore
probably a lower limit, due to extinction.  Even so, the inferred
star formation rate is well within the range observed for GRB host
galaxies \citep{djorgovski01, berger01}.  Thus, except for the high
radio flux, the intrinsic properties of the host galaxy are quite
similar to those of GRB afterglow host galaxies discovered to date.
If the OT is a GRB afterglow, it would mark the first detection of
a GRB afterglow in a galaxy with an AGN.

\section{Discussion\label{discussion}}
Detailed theoretical discussions of the implications of the OT and GRB 
hypothesis and discussions of GRB beaming may be found elsewhere 
\citep{moderski00,granot02}.
Here we consider some of the observational consequences.  In two nights
the spectral slope of the OT changed from $-0.92$ to $-1.29$, which is
about a 7$\sigma$ difference. 
These values do not change significantly when the
$u$-band magnitudes, which are the most uncertain, are excluded.
The OT faded modestly, if
at all, at the longest wavelengths observed (i.e., in the $z$ band).
The modest decline rate can be explained by either a late phase ($t \gg
10$~d) of a power-law flux time decay, or by an early ``plateau'' or
slowly declining phase before the onset of a power-law decay, similar
to that of GRB~970508 \citep{galama98,pedersen98} and GRB~000301C
\citep{masetti00}.  During its plateau phase, GRB~970508 reddened, while 
GRB~000301C's plateau was achromatic.  ROTSE-I archival 
images \citep{kehoe01} from 50,
30, 10, and 4 days before and 1, 2, and 3 days after run 745 reveal no
source to a limiting magnitude of $r^{*}\sim 16.0$ ($\sim 15.0$ for
8, 17, and 50 days later), casting doubt on the late phase scenario.
The absolute magnitude of the OT is near the average value reached by
fading afterglows in a only few days \citep{simon01}.  Additionally,
a comparable change in spectral slope was observed for the afterglow
of GRB~970508 during an early ``plateau'' phase \citep{galama98}.
Finally, according to the standard ultrarelativistic external
shock model of GRB afterglows, the slope of the optical spectrum
is expected to steepen by $\approx 1/2$ when the so-called ``cooling
break'' moves through the optical band \citep{sari98}.
If we interpret the observed  steepening of the optical spectrum
as due to this, it implies that the first image was taken $\sim 1$
day after the GRB (see also Granot et al. 2002).  Thus, the OT was
likely imaged in the SDSS during an early phase.

The possible detection of a GRB afterglow in the first
$\approx1500~{\rm deg}^{2}$ of the SDSS favors strong gamma-ray beaming
relative to the optical emission.  We estimate the number of afterglows
that should be seen thus far in the SDSS search by integrating the
inferred GRB rate in the cosmological volume \citep{schmidt01} in which
an afterglow would be detected, to $z\approx 0.5$, which is the rough
spectroscopic limit of $L_\ast$ galaxies by the SDSS.  Using the inferred
total GRB progenitor event rate of $250\rm\ Gpc^{-3}\ yr^{-1}$ at $z=0$
\citep{frail01}, a GRB rate proportional to the star formation rate,
and assuming that GRB afterglows remain 
brighter than $r^* \sim 19$ (the limiting selection magnitude of
these particular SDSS images) for about one week,
the expected number of bursts observed in the $1500~{\rm deg}^{2}$
is $\sim 5 f_{\rm opt}$, where $f_{\rm opt}$ is the fraction of the
sky illuminated by the optical emission.  Frail et al.\ find that
the beaming fraction of the conical fireball producing the GRB is
$f_\gamma \approx 500^{-1}$. If the associated optical display is
confined to the same jet, i.e. $f_{\rm opt} = f_\gamma$, then the
number of afterglow events expected in this SDSS volume is $\sim
10^{-2}$.  Therefore, the observation of this event favors a 
frequency-dependent beaming angle with $f_{\rm opt}
\gg f_\gamma$ (and hence disfavors models such as that of
\citet{dalal02}).  The only relatively close trigger in the current BATSE
archive\footnote{http://gammaray.msfc.nasa.gov/batse/grb/catalog/current/}
in the months prior to the SDSS imaging is GRB990308, which was
identified with an afterglow more than 12~deg away \citep{schaefer99}.
There is no related IPN trigger, and the nearest Ulysses trigger
is an unlocalized event on 1999 March 16 \citep{hurley99}.  
The lack of any clearly
associated GRB is not surprising given the implied beaming fraction,
and may suggest that the OT is a so-called ``orphan afterglow.''

\section{Conclusions\label{conclusions}}
We have discovered an unusual transient object associated with a
galaxy at redshift of $z=0.385$.  No known type of object is entirely
consistent with the observations.  The coincidence with an apparently
normal galaxy, extreme variability, power-law SED, and high luminosity
are suggestive of a GRB afterglow, although other explanations such
as a highly unusual AGN are possible.  If the object is indeed an
afterglow, then the gamma-rays must be strongly beamed relative to
the optical emission.  Further observations of the galaxy or the
discovery of additional transients with similar properties will help
to clarify the nature of this unusual object.

\section{Note Added After Submission\label{note}}

As suggested above, follow-up observations of the OT have been obtained by 
another group \citep{galyam02} since the submission of this paper.  
These new observations indicate that the AGN hypothesis is more likely.  
\citet{galyam02} also find past evidence for variability in the Digitized Sky 
Survey magnitude-calibrated catalog data (only the digitized image data is 
publicly avilable).  The object, if indeed an AGN, is highly unusual and 
further follow-up work is warranted.
The fact that our extensive search for orphan afterglows found only
one candidate establishes empirically that the background rate, while
possibly non-zero, is low.  Thus, although follow-up observations will
be necessary in order to confirm orphan afterglow candidates, the
number of such candidates will be small and easily manageable.  Further
color-selected searches for orphan afterglows employing deep and broad
sky coverage are therefore practicable.

\acknowledgments

The Sloan Digital Sky Survey (SDSS) is a joint project of The
University of Chicago, Fermilab, the Institute for Advanced Study,
the Japan Participation Group, The Johns Hopkins University, the
Max-Planck-Institute for Astronomy (MPIA), the Max-Planck-Institute
for Astrophysics (MPA), New Mexico State University, Princeton
University, the United States Naval Observatory, and the University
of Washington. Apache Point Observatory, site of the SDSS telescopes,
is operated by the Astrophysical Research Consortium (ARC).
Funding for the project has been provided by the Alfred P. Sloan
Foundation, the SDSS member institutions, the National Aeronautics
and Space Administration, the National Science Foundation, the
U.S. Department of Energy, the Japanese Monbukagakusho, and the Max
Planck Society. The SDSS Web site is {\tt http://www.sdss.org/}.
We acknowledge useful discussions with Jean Quashnock and Daniel
Reichart, and thank the ROTSE team for access to their imaging
archive.  JFB and KA were supported in part by NASA (NAG5-10842).

\clearpage

\clearpage
\input{tab1.tex}

\end{document}

%% file: tab1.tex
\begin{deluxetable}{rllccccccc}
\tabletypesize{\footnotesize}
\tablewidth{0pt}
\rotate
\tablecaption{Observations of SDSSp~J124602.52+011318.8 \label{obstable}}
\tablehead{
  \colhead{Run or} &
  \colhead{Local} &
  \colhead{} &
  \colhead{RA} &
  \colhead{DEC} &
  \colhead{} &
  \colhead{} &
  \colhead{} &
  \colhead{} &
  \colhead{} \\
  \colhead{Plate} &
  \colhead{Date} &
  \colhead{MJD} &
  \colhead{Sec.\ Only} &
  \colhead{$\arcsec$ Only} &
  \colhead{$u^*$} &
  \colhead{$g^*$} &
  \colhead{$r^*$} &
  \colhead{$i^*$} &
  \colhead{$z^*$}
}
\tablecolumns{10}
\startdata
\cutinhead{Imaging}
 745 & 1999/03/20 & 51257.3320 & 2.539 & 18.81 & 17.60 $\pm$ 0.01 &
 17.26 $\pm$ 0.02 & 16.92 $\pm$ 0.03 & 16.68 $\pm$ 0.02 &
 16.47 $\pm$ 0.02 \\ 
 756 & 1999/03/22 & 51259.3266 & 2.539 & 18.80 & 17.91 $\pm$ 0.03 &
  17.40 $\pm$ 0.01 & 16.98 $\pm$ 0.02 & 16.70 $\pm$ 0.01 &
  16.41 $\pm$ 0.02 \\
1462 & 2000/05/05 & 51669.3173 & 2.543 & 18.85 & 21.11 $\pm$ 0.11 &
  20.36 $\pm$ 0.03 & 19.42 $\pm$ 0.03 & 18.88 $\pm$ 0.02 &
  18.44 $\pm$ 0.04 \\
\cutinhead{Spectroscopy}
 291 & 2000/04/26 & 51660.26 & \nodata & \nodata & \nodata &
   $21.17 \pm 0.20$ & $19.76 \pm 0.15$ & $19.22 \pm 0.18$ & \nodata \\
 291 & 2001/01/19 & 51928.52 & \nodata & \nodata & \nodata &
   $20.31 \pm 0.37$ & $ 19.47 \pm 0.34$ & $19.08 \pm 0.35$ & \nodata \\
\cutinhead{Galactic Reddening}
\nodata & \nodata & \nodata     & \nodata & \nodata & 0.09 & 0.07 &
  0.05 & 0.04 & 0.03 \\ 
\enddata
\end{deluxetable}